\documentstyle[aps,pra,preprint]{revtex}
\begin{document}
\title{Two-dimensional photonic crystal polarizer}
\author{Chun Zhang, Jun Wan, Feng Qiao}
\address{Surface Physics Laboratory (National Key Lab), Fudan University, 
Shanghai 200433,
People's Republic of China}
\author{Jian Zi\cite{byeline1}}
\address{International Center for Materials Physics, Shenyang 110015, 
People's Republic of China\\
and \\
Surface Physics Laboratory (National Key Lab), Fudan University, Shanghai
200433, People's Republic of China\cite{byeline2}}
\date{\today}
\maketitle

\begin{abstract}
A novel polarizer made from two-dimensional photonic bandgap
materials was demonstrated theoretically. This polarizer is fundamentally 
different from the conventional ones. It can function in a wide frequency 
range with high performance and the size can be made very compact,
which renders it usefully as a micropolarizer in microoptics.
\end{abstract}

\newpage\narrowtext

Since the pioneering work of Yablonovitch\cite{yab:87} and John\cite
{joh:87}, photonic bandgap (PBG) materials have generated considerable
attention \cite{jos:93,sou:93,jmo:94,sou:96} from many differents research 
fields. PBG materials are periodically
modulated dielectric composites that have stop bands for
electromagnetic (EM) waves over a certain range of frequencies because of
multiple Bragg scattering\cite{yab:87,joh:87}, analogous to the electronic
band structures in semiconductors. They represent a new class of materials
that are capable of uniquely controlling the flow of EM waves or photons\cite
{joa:95}. The unique optical properties of photonic crystals render the
fabrication of perfect mirrors\cite{fin:98}, high efficient antenna
\cite{bro:93},
thresholdless lasers, novel optical waveguides\cite{mek:96,lin:98},
microcavities\cite{mcc:91,for:97} and many other unique optical devices
possible \cite{jos:93,sou:93,jmo:94,sou:96,joa:95}.

Polarizer is one of the basic elements in optics. One of the 
challenges in
microoptics and microstructures optics is the fabrication of clever optical
elements and devices, such as refractive, diffractive lenses, gratings, and
polarizers, with compact sizes and high performance. In this paper we show
that two-dimensional (2D) PBG materials can be used to make polarizer. 

Basically, there were three kinds of conventional ways to make polarizers by
using the properties of absorption, reflection and refraction. Some
materials like polaroid have all of their long organic molecules oriented in
the same direction. These molecules absorb radiation of one polarization,
thus transmitting the orthogonal polarization. The reflection of light from
a surface is polarization and angle dependent. At Brewster's angle, which is
material and wavelength dependent, the reflectance of $p$-polarized light
becomes zero, and the reflected light is completely $s$-polarized.
Birefringent crystals, calcite is an example, have different indices of
refraction for different polarizations of light. Two rays of orthogonal
polarizations entering the crystal will be refracted at different angles and
therefore separated spatially on leaving the crystal.

The idea to use 2D PBG materials to make polarizer is basically different
from the above mentioned ones. It is based on the special properties of 2D
PBG crystals. Any unpolarized light can be decomposed into two
components: one with electric field parallel to the periodic plane (TE)
and the other one with magnetic filed parallel to the periodic plane (TM).
In 2D PBG crystals, propagations of the TE and TM polarizations can be 
decoupled \cite{note1}.
As a consequence, the TE and TM polarizations have their own band
structures and PBGs. The typical band structures of a 2D PBG crystal are
shown in Fig. \ref{fig1}, from which some general features of propagation of
EM waves can be 
postulated. The transmission of an EM wave is dependent on its
frequency and the band structures of the PBG crystal. In the overlapping region
of TE and TM bands both TE and TM waves can transmit. In the overlapping
region of TE and TM PBGs the propagations of both TE and TH waves are
forbidden. In the overlapping region of the TE (TM) bands and TM (TE) PBG, 
only TE (TM) wave can transmit due to the fact that TM (TE) wave cannot 
propagate in the region of its PBG. As a result, the outgoing wave will have 
only one polarization and is perfectly polarized.

For testing purposes, a 2D PBG crystal consisting of dielectric rods
arranged in the square lattice, shown in Fig. \ref{fig2}, is used to 
make polarizer. 
The lattice constant of the lattice is $a$ and
the radius of the rod is 0.25$a$. The dielectric constant of rods is $%
\epsilon =14.0$. The background is air with $\epsilon _b=1.0$. The
calculated projections of photonic band structures in real space 
for this 2D square PBG crystal is shown
in Fig. \ref{fig3}. The band structures of TE and TM waves are calculated 
by using the plane wave expansion method. Owing to the
introduction of a periodicity in 2D, the wavevector will be limited to 
$\pi /a$. A large PBG is opened for the TM wave
in the reduced frequency range from 0.2 to 0.34 due to the spatially
periodic modulation of dielectric constants. From the above discussions, for
an incident EM wave with reduced frequency ranging from 0.2 to 0.34,
the transmission of the TM component is hence forbidden
and that of the TE component is allowable. As a result, the outgoing light
will have only TE component. The frequency range from 0.2 to 0.34 is the
working window if this structure is used to make polarizer.

The performance of a polarizer is conventionally characterized by the degree
of polarization and transmittance. The transmission is calculated by the
transfer matrix method \cite{pen:92}.

The degree of polarization $P$ is defined by 

\begin{equation}
P=\frac{\left| I_{\text{TE}}-I_{\text{TM}}\right| }{I_{\text{TE}}+I_{\text{TM%
}}}
\end{equation}
where $I_{\text TE}$ ($I_{\text TM}$) is the intensity of the outgoing TE (TM)
component. For a natural light, $P=0$ and for a complete polarized light, $%
P=1 $. The transmittance $T$ of a polarizer is defined here as the ratio of
the intensity of the TE wave passing through a polarizer to the incident
intensity of the TE wave

\begin{equation}
T=\frac{I_{\text{TE}}(\text{out})}{I_{\text{TE}}(\text{in})}
\end{equation}
where in and out stand for the incident and outgoing waves, respectively.
For a perfect polarizer, $T=1$ is expected.

The generic 2D PBG polarizer consists of eight layers of dielectric rods
along the $y$ direction. The incident light is also along the $y$ direction.
To check the performance of the polarizer, we display the calculated degree
of polarization $P$ and transmittance $T$ in Fig. \ref{fig4}. 
Within the range of the
reduced frequency from 0.2 to 0.34 the degree of polarization $P$ is almost
1, indicated that this polarizer is excellent in this frequency range.
The transmittance of this polarizer is also very large. 

This kind of polarizer possesses the other virtues. Because of the scale
invariance of PBG materials, only by adjusting the spatial period $a$, we
can make the polarizer adaptable for the desired range of frequency and at
the mean time the degree of polarization $P$ and the transmittance $T$ are
the same. For example, for $a=1$ $\mu $m, the working frequency window is 6
to 10 THz; for $a=10$ $\mu $m, the window is 0.6 to 1 THz; for $a=1$ mm, the
windows is 60 to 100 GHz. The gap to midgap frequency ratio is rather
larger, close to 50 \%.

The miniaturization of a polarizer is rather difficult to achieve in the
conventional polarizers. The 2D PBG polarizer proposed here, however, can be
made with rather small size, which may have potential use in microoptics.
By adjusting the period parameter $a$, we could obtain a polarizer working in
the desired frequency range. There features are absent in the conventional
polarizers.

We denmenstrate in this paper the possibility to make polarizer from 2D PBG
materials. The example given here is the square lattice consisting of 
2D dielectric rods. It can be easily realized in the range of millimeter 
waves and microwaves. It should be noted the air rod structures with dielectric
background may be more amiable to realize in the optical and IR wavelengths. 

{\bf Acknowledgments}: We thank W. Lu and H. Chen for interesting
discussions. This work was supported by the National Natural Science
Fundation of China under Contract No. 69625609.

\begin{figure}[tbp]
\caption{Typical photinic band structures of a 2D PBG crystal. The TE and TM
waves are decoupled in 2D PBG crystals. The hatched areas stand for the
photonic bands. Between bands there exists PBGs.}
\label{fig1}
\end{figure}

\begin{figure}[tbp]
\caption{Schematic top view of a 2D PBG crystal.}
\label{fig2}
\end{figure}

\begin{figure}[tbp]
\caption{Calculated band structures of the 2D PBG crystal. The
frequency $\omega$ is in the units of $c/2\pi a$, where $a$ is the period
and $c$ is the speed of light in vacuum. The frequency can be tuned by
adjusting the parameter $a$.}
\label{fig3}
\end{figure}

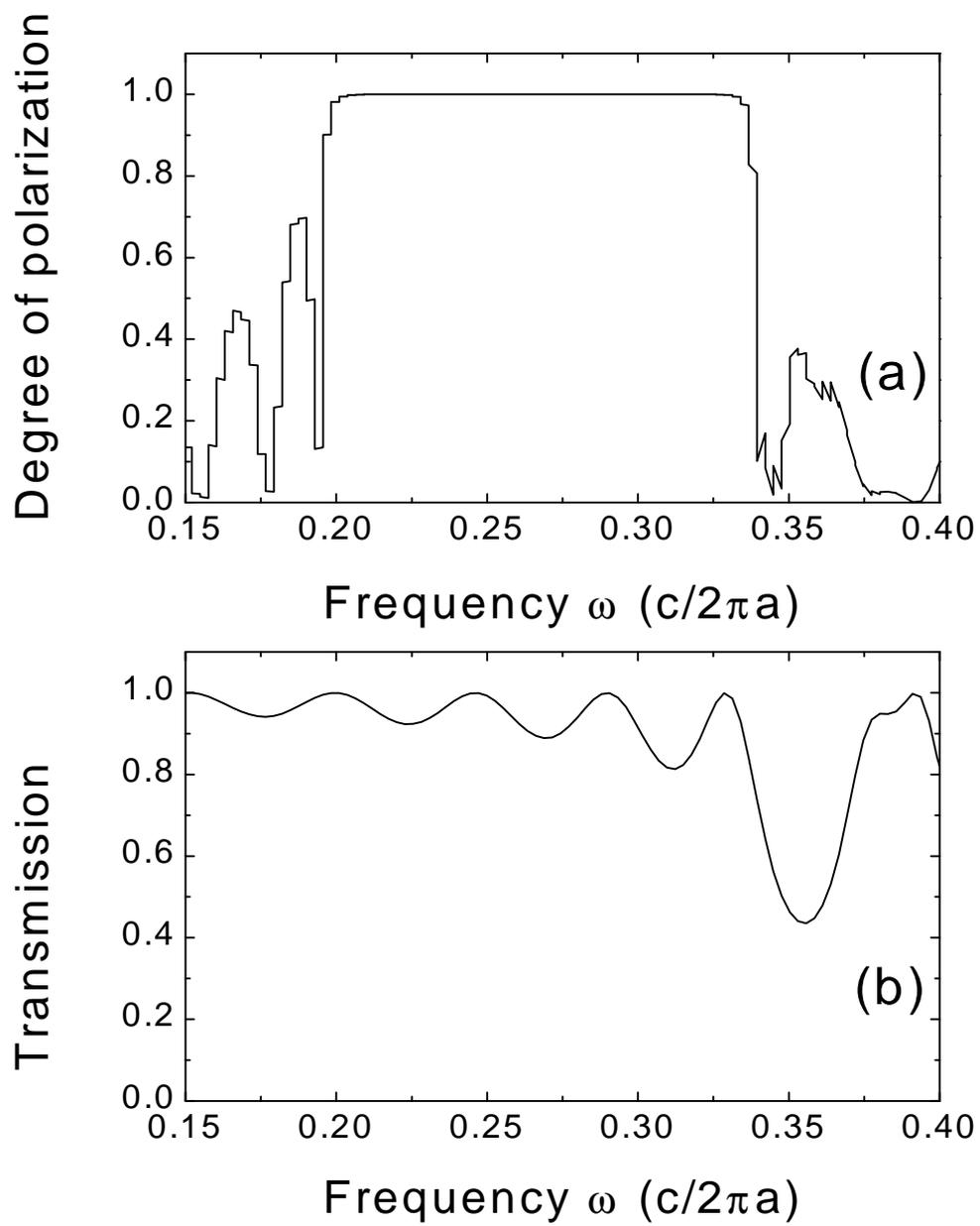
\begin{figure}[tbp]
\caption{ Calculated degree of polarization (a) and transmittance (b)
of the 2D PBG polarizer.}
\label{fig4}
\end{figure}


\begin{references}
\bibitem{note1}  In three dimensions, there is no decoupling of the two
polarizations.
\end{references}
\end{document}